\documentclass{PoS}

\PoS{PoS(LAT2005)064}

     \title{Anisotropic lattice QCD study of pentaquark baryons in spin 3/2 channel }

\ShortTitle{Anisotropic lattice QCD study of pentaquark baryons in spin 3/2 channel }

\author{\speaker{Takumi Doi}
        \thanks{The Monte Carlo simulations have been performed on the
	  NEC SX-5 supercomputer at Osaka University.}\\
        RIKEN BNL Research Center, Brookhaven National Laboratory, Upton, New York 11973, USA\\
        E-mail: \email{doi@quark.phy.bnl.gov}}

\author{Noriyoshi Ishii\\
        Department of Physics, H-27, Tokyo Institute of Technology, \\
	2-12-1 Oh-okayama, Meguro, Tokyo 152-8551, Japan\\
       E-mail: \email{ishii@rarfaxp.riken.jp}}

\author{Yukio Nemoto\\
        Department of Physics, Nagoya University, Furo, Chikusa, Nagoya 464-8602, Japan\\
        E-mail: \email{nemoto@hken.phys.nagoya-u.ac.jp}}

\author{Makoto Oka\\
        Department of Physics, H-27, Tokyo Institute of Technology, \\
	2-12-1 Oh-okayama, Meguro, Tokyo 152-8551, Japan\\
        E-mail: \email{oka@th.phys.titech.ac.jp}}

\author{Hideo Suganuma\\
        Department of Physics, Kyoto University, Kitashirakawaoiwake, Kyoto 606-8502, Japan\\
        E-mail: \email{suganuma@th.phys.titech.ac.jp}}

\abstract{
We perform the comprehensive analysis of
the pentaquark (5Q)  in $J^P=3/2^\pm$ channel
using anisotropic quenched lattice QCD.
We employ the standard Wilson gauge action  at $\beta=5.75$  
and  the  $O(a)$ improved  Wilson (clover)  quark
action  on a  $12^3 \times 96$ lattice
with the  renormalized anisotropy  as $a_{\rm s}/a_{\rm t} =  4$.
A large number of gauge configurations as $N_{\rm conf}=1000$ is
analyzed, which is found to be essential to achieve a reliable measurement.
We study three types of the 
Rarita-Schwinger  intepolationg fields with $I=0$:
(a)  the  NK$^*$-type,  (b) the  (color-)twisted NK$^*$-type, (c) a diquark-type.
As a result, we find only massive states as
$m_{\rm 5Q} \simeq 2.1-2.2$ GeV in $J^P=3/2^-$ channel, 
and $m_{\rm 5Q} = 2.4-2.6$ GeV in $J^P=3/2^+$ channel
in the chiral limit.
The analysis with the hybrid boundary condition (HBC)
is performed 
to distinguish whether these resonances are compact 5Q resonances
or two-particle scattering states.
No low-lying compact 5Q resonance states are found below 2.1GeV.
}

\FullConference{XXIIIrd International Symposium on Lattice Field Theory\\
                 25-30 July 2005\\
                 Trinity College, Dublin, Ireland}

\newcommand{\Ref}[1]{Ref.\protect\cite{#1}}

\newcommand{\Fig}[1]{Fig.~\protect\ref{#1}}
\newcommand{\Figs}[1]{Figs.~\protect\ref{#1}}

\newlength{\Tatescale}
\newcommand{\zr}[1]{\mbox{\hspace*{#1em}}}

\newcommand{\ZZ}{\mbox{\sf Z\zr{-0.45}Z}}

\newenvironment{@@@}{\mbox{ }\\{\bf @@@ begin @@@}\\}{\mbox{ }\\{\bf @@@ end @@@}\\}

\newlength{\figwidth}
\newcounter{subfigure}
\newcommand{\Cut}[1]{}

\makeatletter
\def\simleq{\mathrel{\mathpalette\gl@align<}}
\def\simgeq{\mathrel{\mathpalette\gl@align>}}
\def\gl@align#1#2{\lower.6ex\vbox{\baselineskip\z@skip\lineskip\z@
     \ialign{$\m@th#1\hfill##\hfil$\crcr#2\crcr\sim\crcr}}}
\makeatother
\newcommand{\zi}{$\Theta^+$}
\newcommand{\alt}{\simleq}

\begin{document}

\section{Introduction}

The announcement of the discovery of 
the new-particle $\Theta^+$ (1540) 
with narrow width 
by the LEPS group at SPring-8\cite{nakano}
has triggered enormous studies to
understand this mysterious state.
In fact, $\Theta^+$ has 
baryon number $B=+1$, charge $Q=+1$
and strangeness $S=+1$,
and therefore has the minimal configuration of
$uudd\bar{s}$, which is manifestly exotic.
Although the impact of this discovery 
is undoubted, the experimental status 
is quite controversial: 
several groups confirmed
the existence of \zi\cite{hicks}, while others reported
the null results\cite{hicks}.
Note that even positive results
have not determined the quantum numbers
such as spin and parity.
%
%
Numerous theoretical studies have been performed
as well to understand the nature of \zi\cite{oka}.
In particular,
one of the central issues
is to realize the mechanism for the extremely 
narrow width 
as $\Gamma \simleq 1$MeV.
Among several scenarios proposed so far, 
$J^P=3/2^-$ possibility\cite{hosaka} is interesting to investigate.
In fact, 
if \zi is $J^P=3/2^-$ state,
the decay to the KN scattering state
becomes the d-wave.
Because it is expected that  the special configuration $(0s)^5$ is  dominant in the
ground-state in $J^P=3/2^-$ channel,
the  decay to the d-wave is suppressed by the wave function factor.
Note that there is further  suppression by the d-wave centrifugal barrier,
leading  to  the  significantly  narrow  decay  width.
However, 
this mechanism has 
a possible disadvantage 
that such a state
tends  to be  massive due  to  the color-magnetic  interaction in  the
constituent quark models.
%
$J^P=3/2^+$ possibility attracts another interest considering 
the diquark model\cite{jaffe}. In this model, 
$J^P=1/2^+$ is predicted for $\Theta^+$, where 
$\Theta^+$ is composed
of an anti s quark and
two scalar diquarks with relative p-wave angular momentum.
Narrow width 
is expected from this exotic structure
as well as the p-wave centrifugal barrier.
If QCD really exhibits this scenario, 
we expect that $J^P=3/2^+$ state also exists
as a LS-partner of $\Theta^+$.
%
%
%
%
Under these circumstances,
we present anisotropic lattice QCD results on 5Q states
in $J^P=3/2^\pm$ channels.
Note that 
while 
there  have been  several lattice  QCD  calculations of  5Q states%
\cite{scikor12,sasaki,chiu,kentacky,ishii12,rabbit,
lasscock12,alexandrou12,csikor122,holland,lasscock32}, these
studies  are restricted to  $J^P=1/2^\pm$ channels  except for  a very
recent one \cite{lasscock32}.
%


\section{Formalism}

We study the pentaquark (5Q) state analyzing the 
Euclidean two-point correlator as
$  G_{\mu\nu}(\tau)
  \equiv
  \sum_{\vec x}
  \left\langle
  \psi_{\mu}(\tau,\vec x) \bar\psi_{\nu}(0,\vec 0)
  \right\rangle 
  \label{correlator}$,
where $\psi_\mu$ is a Rarita-Schwinger interpolating field.
In principle, we can use any field $\psi_\mu$ with 
appropriate quantum numbers. In practice, however,
some fields may couple strongly to a genuine 5Q state,
while other fields may couple weakly. 
Unfortunately, there is no empirical knowledge 
about the structure of the 5Q state, i.e., 
about the suitable interpolating field,
we perform the comprehensive study 
using the three types of fields
and examine  how the results
depend on the choice of the fields.
We employ following isoscalar  fields:
(a) NK$^*$-type,
$
  \psi_{\mu}
  \equiv
  \epsilon_{abc}
  \left( u^T_a C\gamma_5 d_b\right) u_c
  \cdot
  \left( \bar{s}_d \gamma_{\mu} d_d \right)
  + (u\leftrightarrow d)
$
(b) {(color-)twisted  NK$^*$-type},
which is an extention to the one in Ref.\cite{scikor12},
$
  \psi_{\mu}
  \equiv
  \epsilon_{abc}
  \left( u^T_a C\gamma_5 d_b\right) u_d
  \cdot
  \left( \bar{s}_d \gamma_{\mu} d_c \right)
  + (u\leftrightarrow d)
$
(c) diquark-type\cite{sasaki},
which is an extention to the one in Ref.\cite{sugiyama},
$
  \psi_{\mu}
  \equiv
  \epsilon_{abc}
  \epsilon_{def}
  \epsilon_{cfg}
  \left( u_a^T C\gamma_5 d_b \right)
  \left( u_d^T C\gamma_5 \gamma_{\mu} d_e \right)
  C\gamma_5 \bar s_g 
$.
%
%
%
%
To analyze spin 3/2 state, 
we decompose 
the correlator as
$
  G_{ij}(\tau)
  =
  {\bf P}^{(3/2)}_{ij} G^{(3/2)}(\tau)
  +
  {\bf P}^{(1/2)}_{ij} G^{(1/2)}(\tau),
  \label{correlator.1}
$
where
$G^{(3/2)}(\tau)$ and $G^{(1/2)}(\tau)$  denote the
spin 3/2  and 1/2  contributions to $G(\tau)$,  respectively, and
$  {\bf P}^{(3/2)}_{ij}
  \equiv
  \delta_{ij}   -  (1/3)\gamma_i\gamma_j,
  {\bf P}^{(1/2)}_{ij}
  \equiv
  (1/3)\gamma_i\gamma_j$.
%
%
%
%
In order to study both parity states,
the parity projection is performed as well
with $P_\pm \equiv (1\mp \gamma_4)/2$.
Note that all of the fields transform as 
$\psi_i(\tau,\vec{x}) \rightarrow -\gamma_4\psi_i(\tau,-\vec{x}) (i=1,2,3)$ 
under the spatial reflection of quark fields as 
$q(\tau,\vec{x}) \rightarrow +\gamma_4 q(\tau,-\vec{x})$.


In the 5Q state study from lattice QCD,
one of the most 
important issues is the discrimination
between two-particle scattering states 
and genuine 5Q compact resonances.
In fact,
in $J=3/2, I=0$ channel,
NK and NK$^*$ scattering states are expected to 
have large contribution 
in the 5Q correlator.
To resolve this problem, we use two
distinct spatial boundary conditions(BC),
i.e., the periodic BC (PBC) 
and the hybrid BC (HBC), 
which is recently proposed in
\Ref{ishii12}.
In PBC, we impose the spatially periodic BC on u,d and s-quarks.  As
a result, all the hadrons are subject to the periodic BC, and 
their momenta are quantized as 
$  p_i = 2 n_i \pi /L, ( n_i \in \ZZ )$,
where $L$ denotes  the spatial extent of the  lattice. 
%
%
%
%
%
On the other hand, in HBC, we impose the spatially anti-periodic BC on
u  and d-quarks,  whereas  the  spatially periodic  BC  is imposed  on
s-quark.      Since    N($uud,udd$),     K($u\bar{s},d\bar{s}$)    and
K$^*$($u\bar{s},d\bar{s}$) contain odd numbers of u and d quarks, they
are subject to the anti-periodic BC
and the momenta are quantized as $  p_i = (2 n_i +1)\pi /L$.
The crucial point is that
the energy of the low-lying two-particle scattering state,
$
  E_{\rm min}
  \simeq
  \sqrt{m_N^2 + |\vec{p}_{\rm min}|^2} + 
  \sqrt{m_{\tilde{K}}^2 + |\vec{p}_{\rm min}|^2} \mbox{($\tilde{K}=K$ or $K^*$)},
$
is different between PBC and HBC
because of the difference of the quantization of momenta.
%
In fact,
a drastic change is expected in the s-wave NK$^*$ state.
%
%
In PBC, the lowest energy of NK$^*$ state is 
$
  E_{\rm min}^{\rm PBC}
  \simeq
  m_N + m_{K^*}.
$
In  contrast, since  both N  and K$^*$  are required  to have
non-vanishing momenta $(\pm \pi/L, \pm \pi/L, \pm \pi/L)$ in HBC,
the lowest energy is raised up as
$
    E_{\rm min}^{\rm HBC}
  \simeq
  \sqrt{m_N^2 + 3\pi^2/L^2} + \sqrt{m_{K^*}^2 + 3\pi^2/L^2}
$,
where the shift of energy amounts typically  to a few hundred MeV for $L\sim
2$ fm.
%
%
The energy shift 
appears
in NK(d-wave), NK(p-wave), NK$^*$(p-wave)  as well.
Note that the minimum momentum in PBC is not
$\vec{p}=\vec{0}$, but 
$\vec{p}=(\pm 2\pi/L,0,0), (0, \pm 2\pi/L,0), ( 0,0,\pm 2\pi/L)$,
because $\vec{p}=\vec{0}$ state can couple
only to s-wave scattering states.
Therefore, 
the lowest energies are expressed
as
$
    E_{\rm min}^{\rm PBC}
  \simeq
  \sqrt{m_N^2 + 4\pi^2/L^2} + \sqrt{m_{\tilde{K}}^2 + 4\pi^2/L^2}
$
and
$
    E_{\rm min}^{\rm HBC}
  \simeq
  \sqrt{m_N^2 + 3\pi^2/L^2} + \sqrt{m_{\tilde{K}}^2 + 3\pi^2/L^2}
$.
%
%
%
These energy difference between PBC and HBC of 
scattering states contrast
strongly with one in a compact 5Q resonance.
In fact, 
because 
$\Theta^+$($uudd\bar{s}$) 
contains even  number of u and  d quarks,
it is  subject  to the  spatially
periodic BC not only in PBC but also in HBC,
which allows $\Theta^+$ to have
$\vec{p}=\vec{0}$ in both of PBC and HBC.
Therefore, 
the energy difference of a compact 5Q state
is expected to be marginal,
since it originates  only from the change  in its  intrinsic structure.
In this way, 
we can identify 
whether the observed states on the lattice are
compact 5Q resonances or scattering states,
by examining the energy difference between
PBC and HBC.

\section{Lattice QCD parameters and Numerical results}

We calculate 5Q correlators
using anisotropic quenched lattice QCD.
To generate gauge field  configurations, 
we  adopt the  standard Wilson  gauge  action at  $\beta=5.75$ on  the
$12^3\times 96$ lattice  with the renormalized anisotropy $a_{s}/a_{t} = 4$.  
Note that the anisotropic lattice is  
suitable for high-precision    measurements    of    temporal    correlators.
The  lattice unit  is determined  as $a_{s}^{-1} = 1.100(6)$ GeV
from the Sommer parameter $r_0^{-1} = 395$ MeV, and thus 
the physical lattice volume amounts to $(2.15\mbox{fm})^3 \times
(4.30\mbox{fm})$.
We use  totally 1000  gauge field  configurations to
achieve the high statistics analysis.
This is quite
essential  for our  study, because  the  5Q correlators  for spin  3/2
states are found to be rather noisy.
%
%
For the quark fields,
we adopt the $O(a)$-improved Wilson (clover) action on the 
anisotropic lattice to suppress the discretization error.
We  adopt four  values  of the  hopping  parameters
as  $\kappa=0.1210 (0.0010) 0.1240$,  
which corresponds to $m_{\pi}/m_{\rho} =
0.81, 0.78, 0.73$ and $0.66$
and roughly covers
the region $m_s \alt m \alt 2 m_s$.
By keeping $\kappa_s = 0.1240$ fixed for s quark,
we change $\kappa=0.1210-0.1240$ for u and d quarks
for chiral extrapolation.
%
%
In order to enhance  the  low-lying spectra,
we  use a  gaussian smeared  source (gaussian size of $\rho\simeq 0.4$ fm)
and point sink correlator in the Coulomb gauge.
For details of the lattice QCD calculations, see Ref.\cite{penta32}.

\subsection{$J^P=3/2^-$ 5Q spectrum}

\begin{figure}[t]
\vspace*{-8mm}
\begin{center}
\begin{tabular}{ccc}
\begin{minipage}{50mm}
\begin{center}
\includegraphics[height=0.8\textwidth,angle=-90]{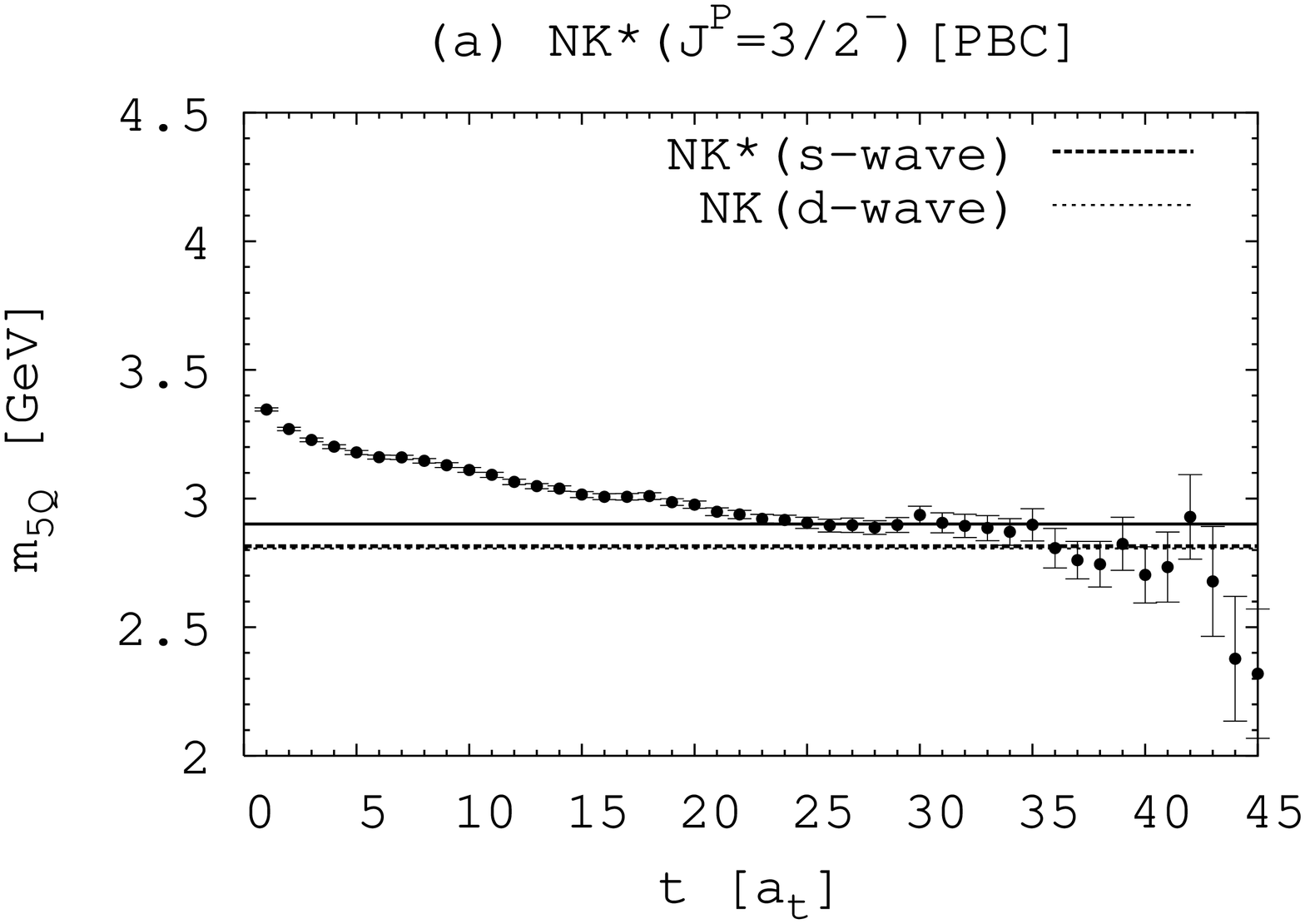}  \\
\includegraphics[height=0.8\textwidth,angle=-90]{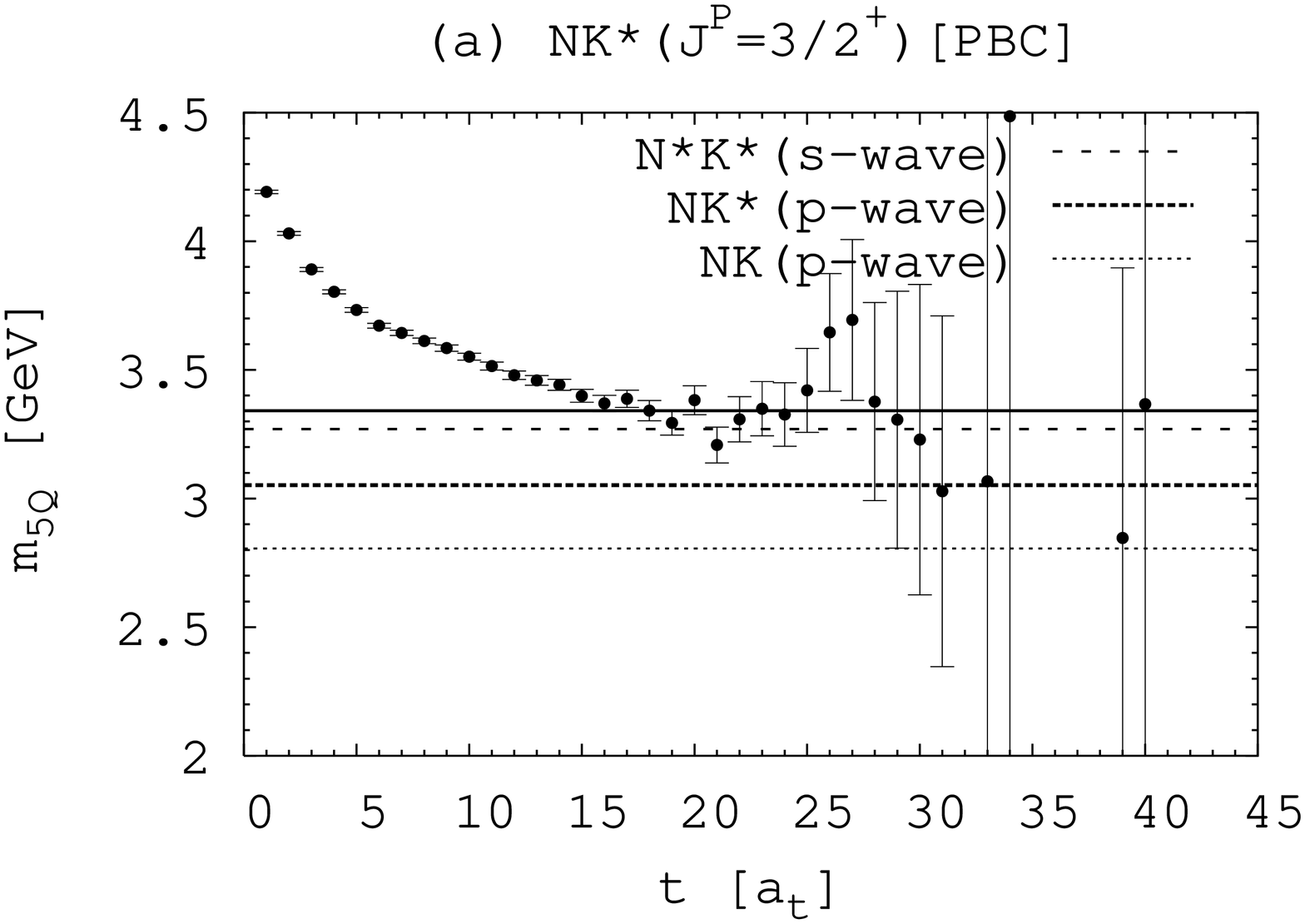}  
\end{center}
\end{minipage}&
\begin{minipage}{50mm}
\begin{center}
\includegraphics[height=0.8\textwidth,angle=-90]{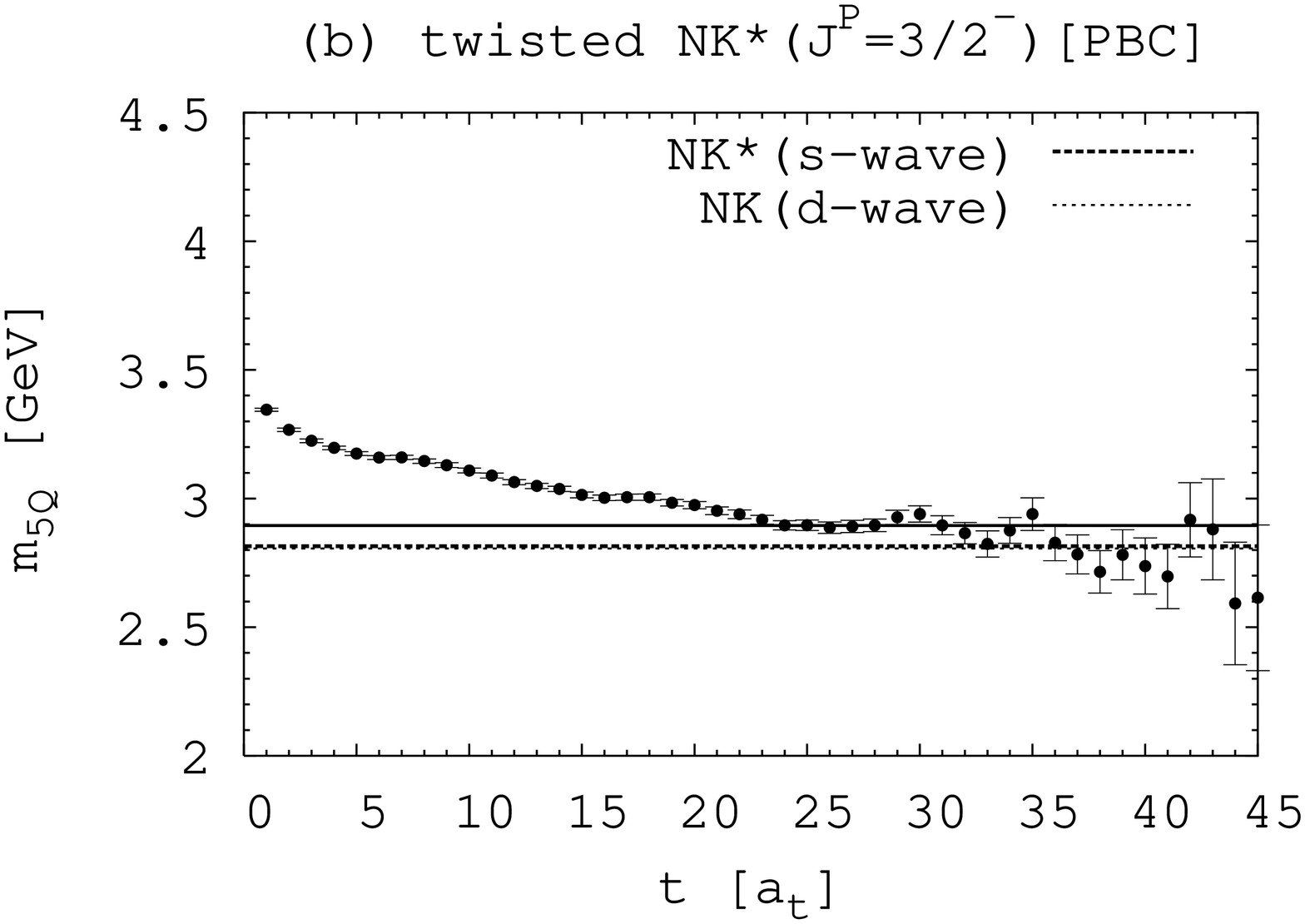}  \\
\includegraphics[height=0.8\textwidth,angle=-90]{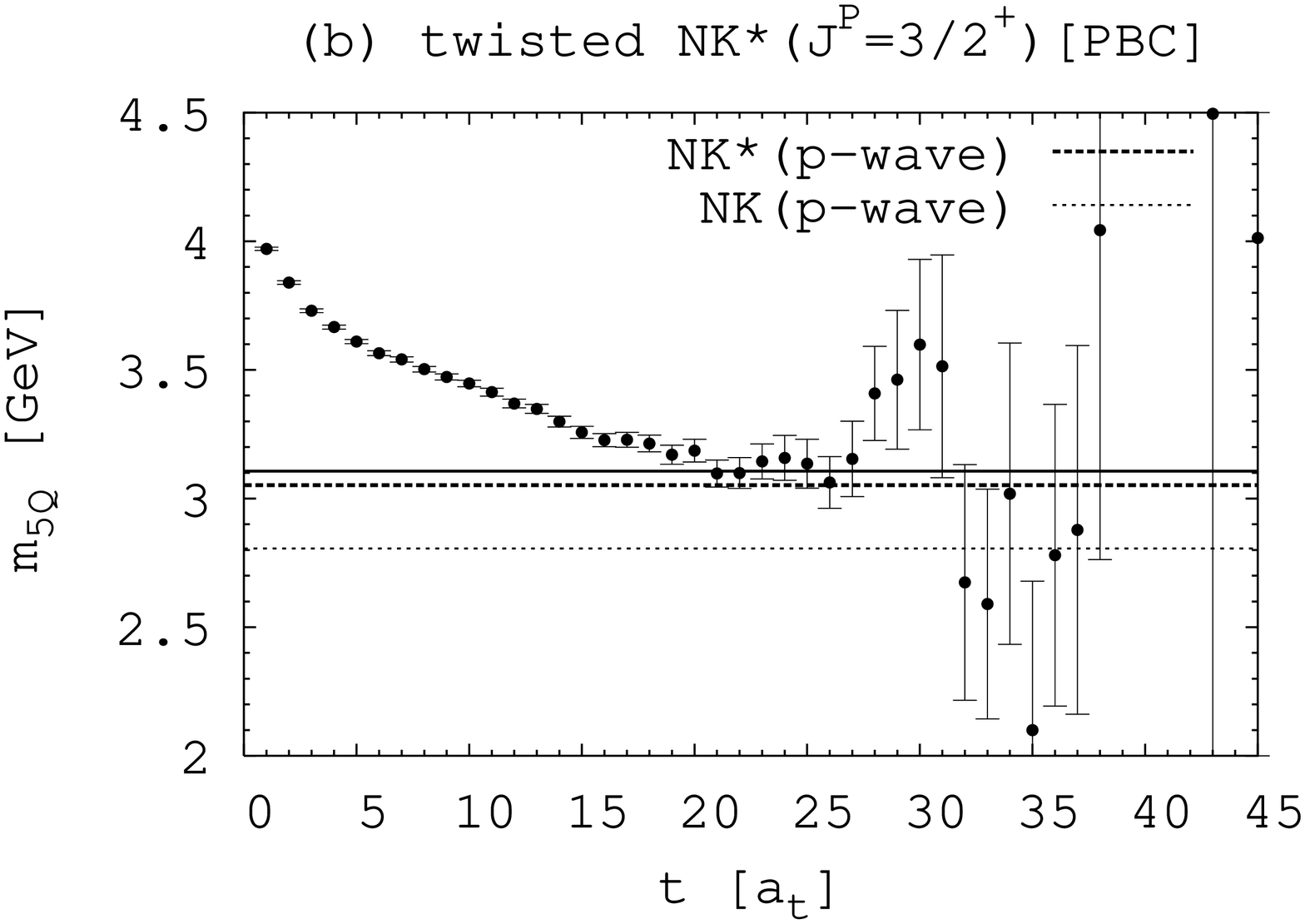}  
\end{center}
\end{minipage}&
\begin{minipage}{50mm}
\begin{center}
\includegraphics[height=0.8\textwidth,angle=-90]{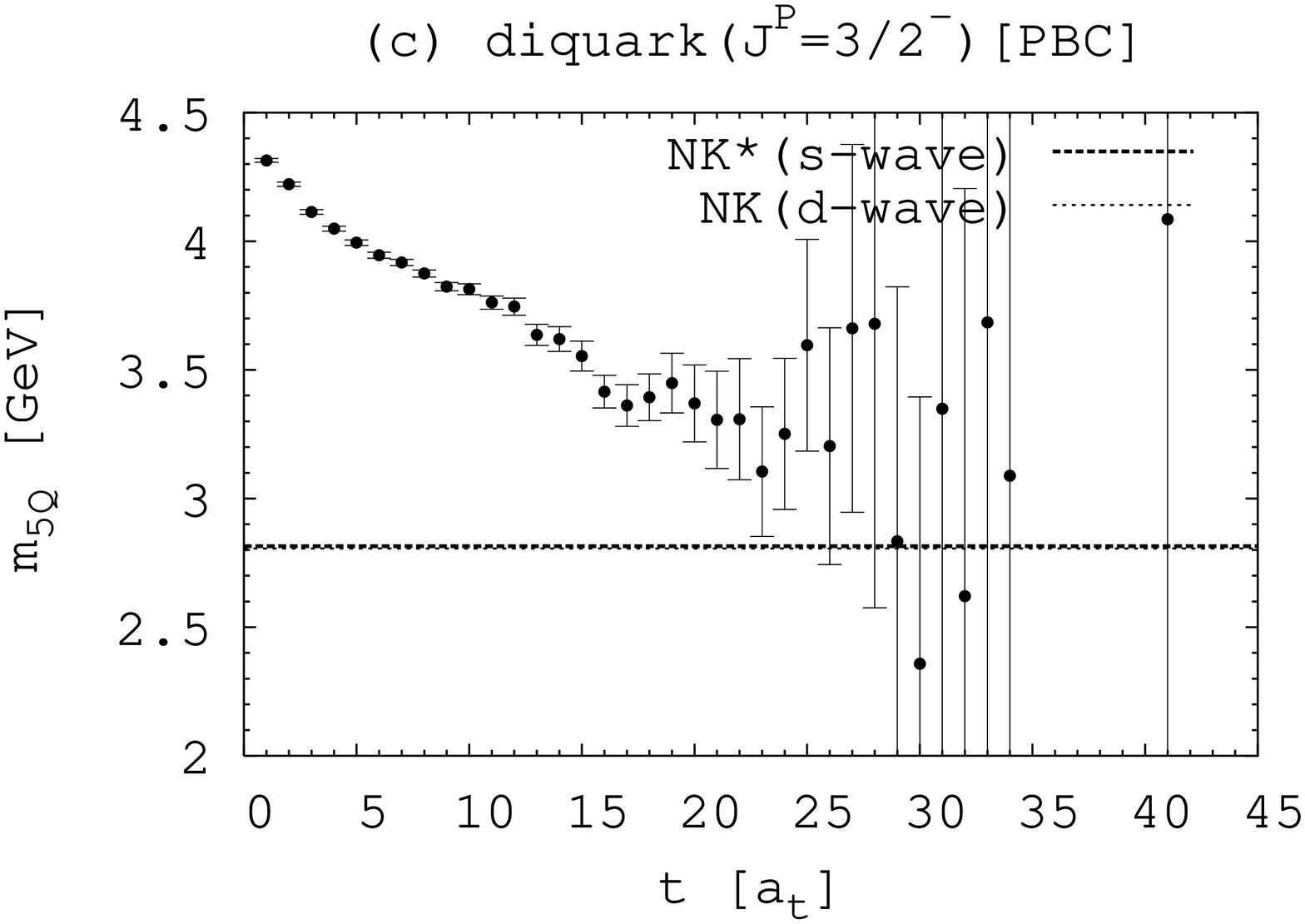} \\
\includegraphics[height=0.8\textwidth,angle=-90]{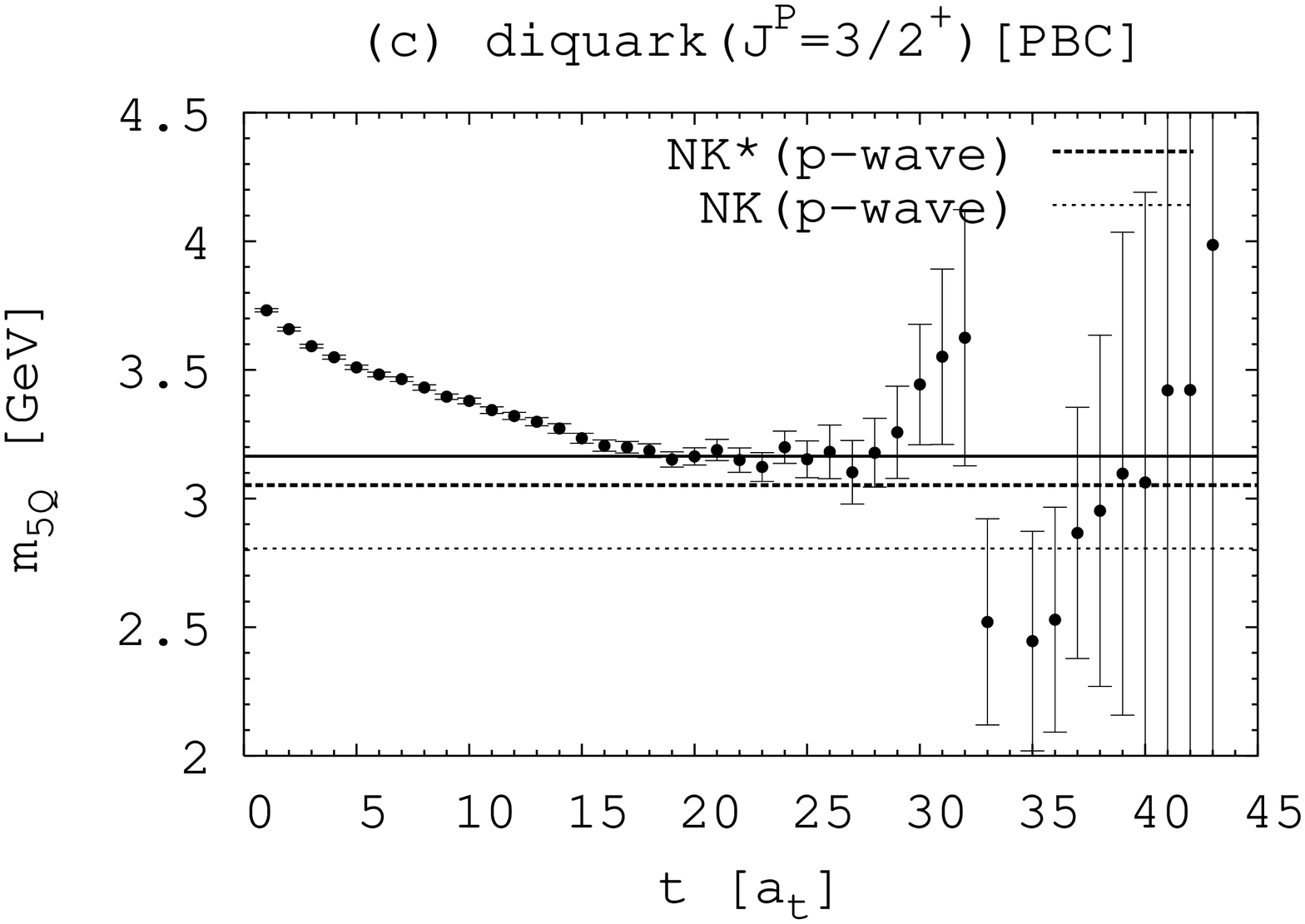}
\end{center}
\end{minipage}
\end{tabular}
\end{center}
\vspace*{-5mm}
\caption{The  5Q effective mass  plots in  $J^P=3/2^-$ (upper) and
$J^P=3/2^+$ (lower) channel  in the
periodic BC(PBC)  for three  types of fields:
(a) the  NK$^*$-type, (b)  the twisted  NK$^*$-type, and  (c) the
diquark-type.
The  dotted  lines  indicate  the  s(p)-wave NK$^*$  and  the  d(p)-wave  NK
threshold for $J^P=3/2^-(3/2^+)$.
The  solid lines  denote  the results  of  the single-exponential  fit
performed in each plateau region.
}
\vspace*{-3mm}
\label{fig.three.half.minus.pbc}
\label{fig.three.half.plus.pbc}
\end{figure}

We consider 5Q spectrum in $J^P=3/2^-$ channel
in PBC.
In the upper side of \Figs{fig.three.half.minus.pbc} (a), (b), (c),
we show the  effective mass  plots 
for  three  
fields, (a)  the
NK$^*$-type, (b) the twisted NK$^*$-type, (c) a diquark-type, respectively.
Unless otherwise indicated, we use $(\kappa_s,\kappa)=(0.1240,0.1220)$
as a typical set of hopping parameters hereafter.
In upper \Figs{fig.three.half.minus.pbc} (a) and (b),
we find the plateaus around
$25  \simleq \tau  \simleq  35$ 
where the  contamination  from excited states 
as well as the backward  propagation
are suppressed.
We perform a single-exponential fit  in this region
and obtain
(a) $m_{\rm 5Q}=2.90(2)$  GeV, 
(b) $m_{\rm 5Q}=2.89(1)$  GeV, respectively.
%
%
On the other hand,
we see that the statistical error is too large 
in upper \Fig{fig.three.half.minus.pbc} (c),
and therefore we do not use 
diquark-type field in 
this channel.
A possible  reason for  such a large  noise is that  
this field does  not  survive in the  non-relativistic
limit.
%
%
%
%
%
Now, we perform the chiral extrapolation.
%
\Fig{fig.three.half.minus.chiral.pbc}   shows   the   5Q   masses   in
$J^P=3/2^-$ channel against $m_{\pi}^2$.  
%
Since the data  behave almost linearly in $m_{\pi}^2$,  we adopt the
linear chiral  extrapolation in $m_{\pi}^2$.
%
As a  result,  we obtain only  massive 5Q
states as  
(a) $m_{\rm 5Q}=2.17(4)$ GeV, 
(b) $m_{\rm 5Q}=2.11(4)$ GeV, respectively,
which is too  heavy to be identified with  
$\Theta^+(1540)$.
Note that 
none of 5Q states appear below the NK  threshold,
although
this threshold
is raised up  by about $200-250$  MeV 
due to  the finite extent of the spatial lattice,
and the 5Q signal is expected to appear  below the
(raised)  NK  threshold  considering  the  empirical  mass  difference
between N+K(1440) and $\Theta^+(1540)$.

To  clarify whether our  5Q states  are compact  resonances or  not, we
perform the analysis with HBC.
With the typical set of hopping parameters, 
it is expected that 
the s-wave NK$^*$ state
is raised  up by $\sim
180$ MeV, the d-wave NK state is lowered down  by $\sim 70$  MeV,
and a compact 5Q state shows no difference
by switching PBC to HBC.
Actual lattice simulations 
show
$m_{\rm 5Q}= 2.98(1)$ GeV
for both of NK$^*$-type and twisted NK$^*$-type fields,
which corresponds to 
upper shift by 80-90MeV.
Although the shift of $m_{\rm 5Q}$ is rather small,
the  value  of $m_{\rm  5Q}$  is  consistent  with the  s-wave  NK$^*$
state
within the statistical error.
Therefore, 
we regard this state as an s-wave NK$^*$ state.
In this way, we do not  observe any
compact 5Q  resonances in  $J^P=3/2^-$ channel below  the raised
s-wave NK$^*$ threshold, i.e., in the region of
$  E
  \alt 
  \sqrt{m_N^2 + \vec p_{\rm min}^2}
  +
  \sqrt{m_{K^*}^2 + \vec p_{\rm min}^2}$, 
with $|\vec p_{\rm min}|\simeq 499$ MeV.

\subsection{$J^P=3/2^+$ 5Q spectrum}

\begin{figure}[t]
\vspace*{-8mm}
\begin{center}
\begin{tabular}{cc}
\begin{minipage}{70mm}
\begin{center}
\includegraphics[height=0.65\textwidth,angle=-90]{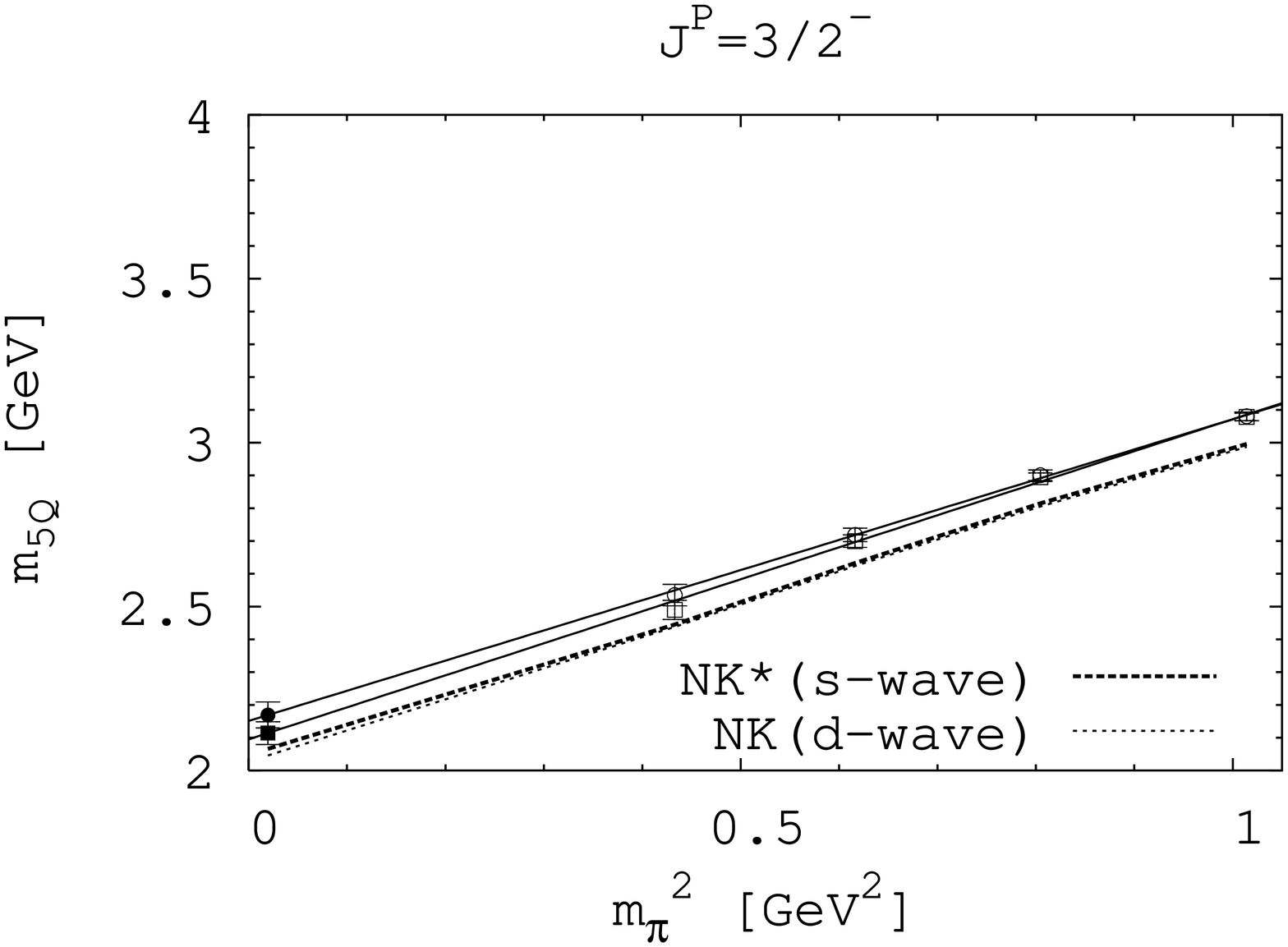}  \\
\end{center}
\end{minipage}&
\begin{minipage}{70mm}
\begin{center}
\includegraphics[height=0.65\textwidth,angle=-90]{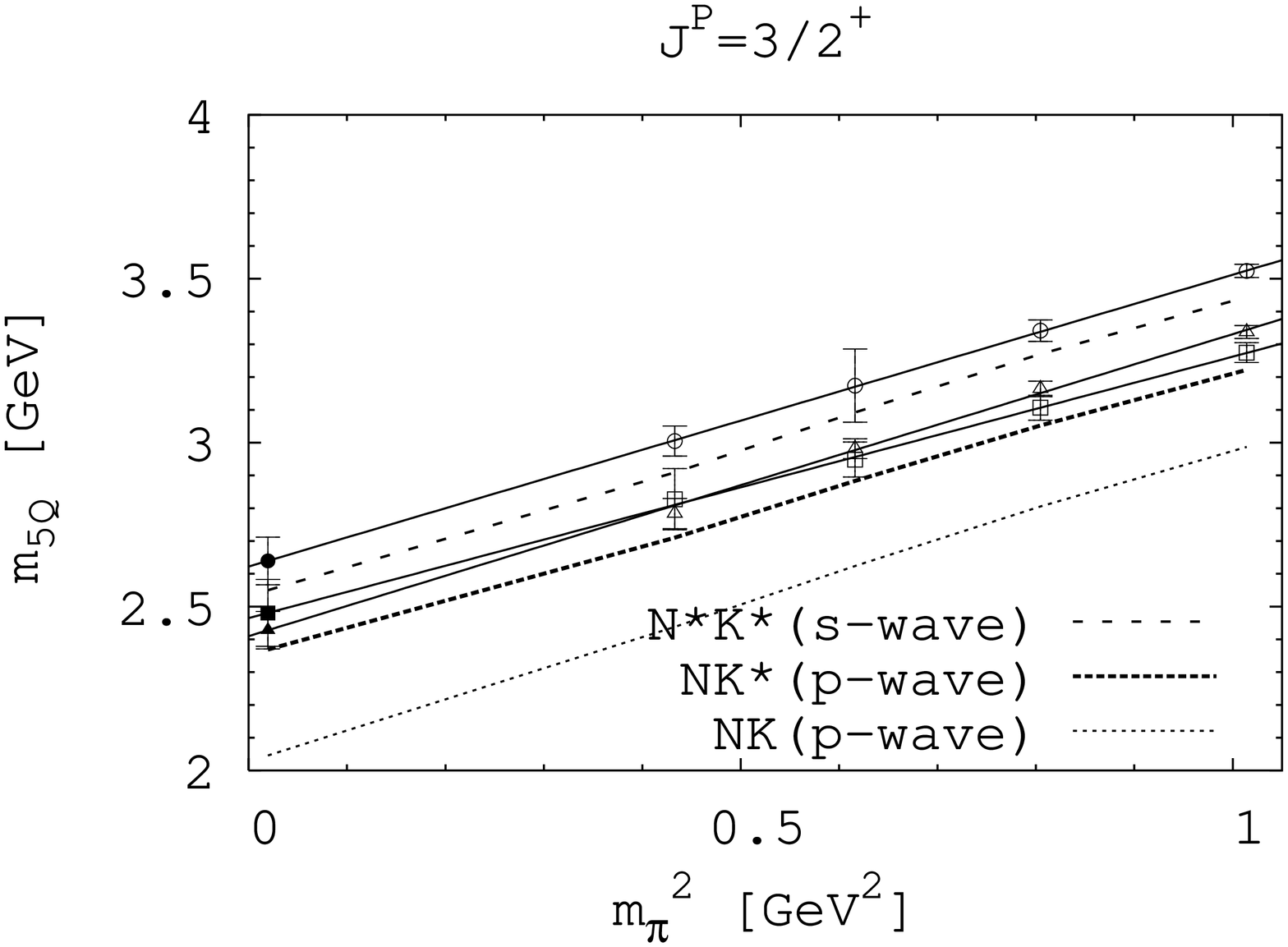}\\
\end{center}
\end{minipage}
\end{tabular}
\end{center}
\vspace*{-5mm}
\caption{$m_{\rm 5Q}$ in  $J^P=3/2^-$ (left) and 
$3/2^+$ 
(right)
channel against  $m_{\pi}^2$ for the
three fields, i.e., (circle) the NK$^*$-type, (box) the
twisted      NK$^*$-type,     and     (triangle)      a     diquark-type.
Data from diquark-type are not plotted in $J^P=3/2^-$ because 
a plateau is not obtained in the effective mass plot.
Open  symbols denote  the direct  lattice results, while  the closed
symbols  and the  solid  lines represent the  results  of the  chiral
extrapolations.
}
\label{fig.three.half.minus.chiral.pbc}
\label{fig.three.half.plus.chiral.pbc}
\vspace*{-3mm}
\end{figure}

We perform the analysis for 5Q spectrum in $J^P=3/2^+$ channel.
The lower side of \Figs{fig.three.half.plus.pbc} (a), (b), (c), show
the  effective mass  plots 
for (a)  the
NK$^*$-type, (b) the twisted NK$^*$-type, (c) a diquark-type,
respectively.
In all figures, we observe the plateaus and 
perform the single-exponential fit.
We obtain
(a) $m_{\rm 5Q}=3.34(3)$ GeV,
(b) $m_{\rm 5Q}=3.11(4)$ GeV and
(c) $m_{\rm 5Q}=3.16(2)$ GeV,
respectively.
Note  that 
(a)
agrees with  the  s-wave N$^*$K$^*$  threshold
while the latter two is very close to NK$^*$ threshold.
Now,  we  perform  the  chiral  extrapolation.
In  \Fig{fig.three.half.plus.chiral.pbc}, $m_{\rm 5Q}$ is  plotted against
$m_{\pi}^2$.  
Using the linear chiral extrapolation in $m_{\pi}^2$,
we   obtain  
(a) $m_{\rm 5Q}=2.64(7)$ GeV,
(b) $m_{\rm 5Q}=2.48(10)$ GeV and
(c) $m_{\rm 5Q}=2.42(6)$ GeV, respectively.
Note that 
we observe  again that 
all  of $m_{\rm 5Q}$ data appear  above the
NK$^*$ p-wave  threshold,  which is  located  above the  
(raised)  NK p-wave threshold.

Next, we perform the HBC analysis.
For $J^P=3/2^+$ channel, the energy shift
is somewhat minor change.
With the typical set of hopping parameters, 
it is expected that 
the s-wave N$^*$K$^*$ state
is raised  up by $\sim
170$ MeV, 
the p-wave NK$^*$ state is lowered down  by $\sim 60$  MeV,
the p-wave NK state is lowered down  by $\sim 70$  MeV
and a compact 5Q state shows no difference
by switching PBC to HBC.
As a lattice result, 
we obtain 
$m_{\rm 5Q}=3.38(2)$ GeV from 
NK$^*$-type field.
Although corresponding shift of 40 MeV
is rather small, 
 $m_{\rm  5Q}$ is again
almost  consistent  with the  s-wave  N$^*$K$^*$  threshold.
Considering  its rather  large  statistical error,  this  5Q state  is
likely to be an s-wave N$^*$K$^*$ state.
The results from other two fields are 
very similar to each other.
We obtain 
(b) $m_{\rm 5Q}=3.02(3)$ GeV
and
(c) $m_{\rm 5Q}=3.08(4)$ GeV,
respectively.
These results correspond to lower shift by 80-90MeV,
which is  considered  to be  consistent  with 
the NK$^*$ p-wave  behavior.
Therefore, these states are likely to be an NK$^*$ p-wave state.
%
%
%
In  this  way,  we observe no signal for compact 5Q resonances.
Although the assignments for plateaus are still afflicted by 
considerable size  of statistical error,
we  can at  least state  that these 5Q  states are  all massive,
which locate above NK$^*$ p-wave threshold
and that
these states are too  heavy to be identified 
with  $\Theta^+(1540)$.

\section{Summary and Conclusions}

We have  studied $J^P=3/2^\pm$ pentaquark(5Q)  baryons in anisotropic
lattice QCD at  the quenched level.
We have employed  the standard Wilson gauge action  
at  $\beta=5.75$
on $12^3\times 96$ lattice with  the renormalized anisotropy
$a_{\rm s}/a_{\rm  t} = 4$.
The large   statistics as $N_{\rm conf}=1000$
has played a 
key role in achieving a solid result in our calculation.
%
For the quark action, we have adopted $O(a)$-improved Wilson (clover) 
action with
the  hopping parameter  as  
$\kappa=0.1210  (0.0010) 0.1240$, 
which  roughly  corresponds to 
$m_s \alt  m \alt  2m_s$. 
We have examined  three types of the interpolating fields as (a)
the  NK$^*$-type,   (b)  the  twisted   NK$^*$-type,  (c)  the
diquark-type.
%
In  $J^P=3/2^-$ channel,  we have observed  plateaus
in the effective mass plots
except for  the diquark-type field.
Employing  the  linear chiral  extrapolations in  $m_{\pi}^2$,
we have obtained $m_{\rm 5Q}\simeq  2.17$ and  $2.11$  GeV  in the chiral limit
for  the
NK$^*$-type and the twisted NK$^*$-type correlators, respectively.
We have performed the HBC analysis and have found that 
both of the observed 5Q  states are s-wave NK$^*$ scattering states.
%
%
In $J^P=3/2^+$ channel,  we have recognized plateaus 
in all the three
effective  mass  plots.   
The chiral extrapolations have lead to 
$m_{\rm 5Q}\simeq 2.64, 2.48, 2.42$  GeV 
for the  NK$^*$-type, twisted NK$^*$-type and 
the diquark-type correlator, respectively.
HBC analyses have been performed and 
the observed 5Q states 
in both of the  twisted  NK$^*$-type   and  the
diquark-type correlators
are most likely to be NK$^*$ p-wave states.
For the NK$^*$-type field, 
it is most
likely to be an s-wave N$^*$K$^*$ state, 
although more statistics  is needed to draw  a definite conclusion.
At any rate, whatever the real  nature of these 5Q states may be, they
are all considerably  massive states in  the physical  quark mass
region in both of $J^P=3/2^\pm$ channels,
and  cannot   be  identified  as  $\Theta^+(1540)$  without
involving a significantly large chiral effect.
For further studies,
it is important to perform the systematic studies of the 5Q states 
using such as 
(1) unquenched full lattice QCD, 
(2) finer and larger volume lattice,  
(3) chiral fermion with small mass,
(4) more sophisticated interpolating field reflecting the structure of 5Q, etc.

\begin{thebibliography}{99}
\bibitem{nakano}
  LEPS Collaboration, T.~Nakano  {\it   et  al}.,
  Phys. Rev. Lett. {\bf 91}, 012002 (2003).

\bibitem{hicks}
  For a review of the experimental status,
  K.H.~Hicks, hep-ex/0504027,
  and references therein.

\bibitem{oka}
  For a review, 
  M.~Oka, Prog. Theor. Phys. {\bf 112}, 1 (2004), and references therein.

\bibitem{hosaka}
  A.~Hosaka,
  Phys. Lett. B{\bf 571}, 55 (2003).


\bibitem{jaffe}
  R.L.~Jaffe and F.~Wilczek,
  Phys. Rev. Lett. {\bf 91}, 232003 (2003).



\bibitem{scikor12}
  F.~Csikor, Z.~Fodor, S.D.~Katz, and T.G.~Kovacs,
  JHEP {\bf 0311}, 070 (2003).
\bibitem{sasaki}
  S.~Sasaki,
  Phys. Rev. Lett. {\bf 93}, 152001 (2004).
\bibitem{chiu}
  T.W.~Chiu and T.H.~Hsieh,
  hep-ph/0403020.
\bibitem{kentacky}
N.~Mathur {\it et al.}, 
  Phys. Rev. D{\bf 70}, 074508 (2004).
\bibitem{ishii12}
  N.~Ishii, T.~Doi, H.~Iida, M.~Oka, F.~Okiharu, and H.~Suganuma,
  Phys. Rev. D{\bf 71}, 034001 (2005).
\bibitem{rabbit}
  T.T.~Takahashi, T.~Umeda, T.~Onogi and T.~Kunihiro, 
  Phys. Rev. D{\bf 71}, 114509 (2005). 
\bibitem{lasscock12}
B.G.~Lasscock {\it et al.}, 
Phys. Rev. D{\bf 72}, 014502 (2005).
\bibitem{alexandrou12}
  C.~Alexandrou and A.~Tsapalis,
  hep-lat/0503013.
\bibitem{csikor122}
  F.~Csikor, Z.~Fodor, S.D.~Katz, T.G.~Kov\'acs, and B.C.~T\'oth,
  hep-lat/0503012.
\bibitem{holland}
  K.~Holland, and K.J.~Juge,
  hep-lat/0504007.
\bibitem{lasscock32}
B.G.~Lasscock {\it et al.},
  hep-lat/0504015.

\bibitem{sugiyama}
  J.~Sugiyama, T.~Doi, and M.~Oka,
  Phys. Lett. B {\bf 581}, 167 (2004).

\bibitem{penta32}
  N.~Ishii, T.~Doi, Y.~Nemoto, M.~Oka and H.~Suganuma,
  hep-lat/0506022, Phys. Rev. D{\bf 72} (2005).


\end{thebibliography}
\end{document}